\documentstyle[aps,prl,twocolumn,epsfig]{revtex}

\def\bra#1{{\langle #1 |}}
\def\ket#1{{| #1 \rangle}}
\def\id{{\hat 1}}
\def\a{{\hat a}}
\def\adag{{\hat a}^\dagger}
\def\H{{\hat H}}
\def\L{{\hat L}}
\def\Ldag{{\hat L}^\dagger}

\begin{document}
\draft

\title{Coupling nanocrystals to a high-$Q$ silica microsphere: \\
entanglement in quantum dots via photon exchange}

\author{Todd A. Brun\thanks{Email:  tbrun@andrew.cmu.edu} \\
Department of Physics, Carnegie Mellon University, \\
Pittsburgh PA 15213 \\
Hailin Wang\thanks{Email:  hailin@oregon.uoregon.edu} \\
Department of Physics, University of Oregon, Eugene OR 97403 }

\maketitle

\begin{abstract}
Coupling nanocrystals (quantum dots) to a high-$Q$ whispering gallery mode
(WGM) of a silica microsphere,
can produce a strong coherent interaction between the WGM and the electronic
states of the dots.  Shifting the
resonance frequencies of the dots, for instance by placing the entire
system in an electric potential, then allows this interaction to be
controlled, permitting entangling interactions between different dots.
Thus, this system could potentially be used to implement
a quantum computer.
\end{abstract}


Recent years have seen a dramatic improvement in the ability to
experimentally manipulate and measure single quantum systems, from
modes of the electromagnetic field to atoms or ions in traps to
increasingly tiny solid-state devices.
These efforts have been further spurred by developments in the field
of quantum computation, where it has been shown that a computer making
full use of quantum mechanics can solve certain problems (such as
factoring and database searching) faster than any known classical
algorithm.  Building such a computer requires well-isolated but
highly controllable quantum systems, with well-defined levels which
can serve as quantum bits (qubits).

A number of proposals have been made for physical implementations of a
quantum computer.  Two of the most promising involved using the internal
electronic states of ions \cite{CiracZoller1995} or atoms
\cite{Pellizzari1995} in a trap as qubits, with single-qubit operations
performed by laser pulses; interactions between qubits would involve an
intermediate ``bus,'' which would be the quantized motion in the case
of ions, an electromagnetic mode of a high-$Q$ cavity in the case of atoms.
In a similar spirit, Sherwin et al. \cite{Sherwin1999}
recently proposed using semiconductor quantum dots (QDs) as qubits, with
interactions mediated by the field of a microcavity.
An earlier proposal by Loss and DiVincenzo \cite{Loss1998} had also
involved QDs, in that case using the spin of a single
electron in a QD, with coupling between neighboring dots
due to the exchange interaction.  Recently, a hybrid of these two
proposals has also been suggested by Imamoglu et al.
\cite{Imamoglu1999}, in which interactions between electron spins of
different dots are mediated by one mode of an optical microcavity.


In this paper we propose and analyze a novel system for deterministic
generation of quantum entanglement of two QDs.  In this system,
semiconductor nanocrystals are attached to the equator of a fused silica
microsphere.  Exchange of photons between the two nanocrystals through
optical interactions between a nanocrystal and a high-$Q$ whispering gallery
mode (WGM) of the microsphere leads to creation of quantum entanglement
between the two nanocrystals.  Entanglement of nanocrystals with slightly
different resonance frequencies can be generated by using an electric field
to tune the relevant
transition frequency of a given nanocrystal to resonate with the WGM.
The ground state and the lowest dark excited state of the nanocrystals serve
as a qubit,
and it is these levels which become entangled.  The proposed scheme also has
the potential to scale up to implement a quantum computer.

Fused silica microspheres are perhaps the best available optical
microresonators \cite{Braginsky1989}.  In these microspheres, WGMs form via
total internal
reflection along the curved boundary.  Slight deformation of typical
microspheres also removes azimuthal degeneracy of the WGMs.  Lowest order
WGMs are thus a ring along the equator of the microsphere.  For a fused
silica sphere with a diameter of 20 $\mu$m, the very small mode volume
leads to
a vacuum electric field (or field strength per photon) of order 150 V/cm at
the sphere surface ($\lambda \approx 600$ nm).
$Q$-factors of these microresonators can
exceed $10^9$, corresponding to a photon storage lifetime near a microsecond and
resulting in highest finesse for optical resonators.  Extremely high
$Q$-factors along with very small mode volume are essential for achieving
strong coherent light-matter coupling, and have made fused silica microspheres
an attractive alternative to the high finesse
Fabry-Perot microresonators currently used in cavity QED studies of atoms.  

Recent experimental attempts to take advantage of WGMs in fused silica
microspheres to achieve strong coherent light-matter coupling have included
putting a fused silica microsphere in atomic vapors \cite{Vernooy1998} or
placing a
semiconductor nanostructure in the evanescent waves of WGMs of a
microsphere \cite{Fan1998}.  Further development, however, has been hindered by 
thermal motion of the atoms in the former case and by significant
$Q$-spoiling occurring at the contact area
in the latter case.  To avoid these problems, we propose to directly
attach semiconductor nanocrystals to the equator of a fused silica
microsphere, as shown in Fig. 1.  

Semiconductor nanocrystals exhibit discrete atomic-like electronic energy
structures \cite{Alivisatos1996}. 
Synthesis of epitaxially grown, nearly defect free, CdSe/CdS
core/shell nanocrystals has been demonstrated recently \cite{Peng1997}.
Near unity quantum
yield has been observed in these nanocrystals even at room temperature.
These nanocrystals can also be covalently bonded to metal or glass surfaces.
Photoluminescence studies of single CdSe nanocrystals revealed extremely narrow
linewidth, limited by instrument resolution \cite{Empedocles1996}.
Extensive experimental and
theoretical studies also indicate that the lowest excited state in these
nanocrystals is a dark state with a lifetime on the order of microseconds
\cite{Efros1996}.
Assuming a radiative lifetime of 10 nanoseconds for a dipole transition in
CdSe nanocrystals, we expect the coherent dipole-coupling rate between the
dipole transition and a resonant WGM with a vacuum field of 150 V/cm to be
$10^9$/sec, much greater than the radiative decoherence rate for the relevant
transition.


In semiconductor nanocrystals, the precise electronic level spacing depends
sensitively on the exact size and shape of the nanocrystals.  In general
two different nanocrystals will have different resonance frequencies
although these resonance frequencies 
can be precisely determined.  This nonuniformity of nanocrystals, however,
can be turned 
into an advantage. If one could
shift the resonance frequencies of all the nanocrystals simultaneously,
it would be possible to bring a single nanocrystal into resonance with 
a cavity mode while the others are all sufficiently detuned to have no
significant interaction with the cavity mode.

One way of producing such a shift is to put the entire system between
two microelectrodes and apply a controlled voltage.  The level spacings
would all be shifted by the quantum confined Stark effect of nanocrystals
\cite{Empedocles1997}.  Provided that the responses
of the various nanocrystals involved have been determined by measurement, it
should be
possible to tune the system to bring any nanocrystal into resonance on demand.
If the voltage changes are slow compared to optical frequencies, this
resonance shift may be treated adiabatically.  (The voltage change may
still be very rapid compared with the timescale of interactions between
the internal state of the nanocrystals and the cavity mode.)

It is also possible to shine a laser beam on individual nanocrystals in order to
drive transitions between their internal levels.  By combining such
laser pulses with carefully timed interactions with the cavity
mode, one can construct a scheme similar to that of Cirac and Zoller
\cite{CiracZoller1995} or Pellizzari et al. \cite{Pellizzari1995}, 
in which laser pulses act as one-qubit ``gates''
while two-bit gates use the photon mode as an intermediary ``bus.''


We now consider two nanocrystals interacting with
a nearly resonant high-$Q$ WGM.
We model the two QDs as three-level systems as shown in Fig. 2, where states
$\ket0$ and
$\ket1$ serve as the logical states of the qubit and $\ket2$ is an
auxiliary state.  Other energy levels are far off-resonance for the relevant
optical interactions and are thus neglected.
The resonance frequency of the $1\rightarrow2$ transition
is assumed to be close but not equal to the frequency of the cavity mode.
This resonance frequency is assumed to differ for the two dots by an
amount $\Delta\omega$.  The entire system is then placed between two
microelectrodes, so that a precisely controlled voltage can be applied,
shifting the frequencies of the two nanocrystals.

The two dots are each coupled to the WGM with a strength $g$.
It is assumed that the timescale $1/g$ is long compared to the optical
timescales.  We change the voltage slowly compared to optical timescales,
but quickly compared to $1/g$; thus, this change can be treated
adiabatically as far as the Hamiltonian is concerned, but as instantaneous
compared to the rate at which photons are emitted or absorbed.

A product state for this model can then be written
\begin{equation}
\ket\Psi = \ket{\phi_A} \otimes \ket{\phi_B} \otimes \ket{\phi_\gamma}\ ,
\end{equation}
where $\ket{\phi_A}$ is the state of dot A, $\ket{\phi_B}$ is the state
of dot B, and $\ket{\phi_\gamma}$ is the state of the cavity mode.
In a rotating frame we can then write the Hamiltonian of this system
\begin{eqnarray}
\H  &=& ig\left( \ket2\bra1 \otimes \id \otimes \a
  - \ket1\bra2 \otimes \id \otimes \adag \right) \nonumber\\
&&  + ig\left(\id \otimes \ket2\bra1 \otimes \a
  - \id \otimes \ket1\bra2 \otimes \adag \right) \nonumber\\
&& + \delta(t) \id \otimes \ket1\bra1 \otimes \id \nonumber\\
&& + (\delta(t) - \Delta\omega) \ket1\bra1 \otimes \id \otimes \id\ .
\label{hamiltonian}
\end{eqnarray}
The function $\delta(t)$ is the detuning produced in the $1\rightarrow2$
resonance due to the time-varying voltage.  Either A or B (or neither)
can be brought into resonance with the cavity mode, but not both at once.
We assume that their frequency difference $\Delta\omega$ is large
compared to $g$.

We begin in a product state
\begin{equation}
\ket{\Psi_i} = {1\over2}\left( \ket0 + \ket1 \right) \otimes
  \left( \ket0 + \ket1 \right) \otimes
  \ket0\ .
\end{equation}
and apply a $\pi$-pulse to the $1\rightarrow2$ transition of dot A.
(The coherent superposition state of each dot can be created by 
using resonant two-photon absorption.)
The voltage is tuned such that $\delta(t)=\Delta\omega$ for a time
$\pi/2g$, then $\delta(t)=0$ for a time $\pi/g$,
then back to $\delta(t)=\Delta\omega$ for another $\pi/2g$.
Finally, a second $\pi$-pulse is applied to dot A.  In the absence of
noise, the final state becomes
\begin{eqnarray}
\ket{\Psi_f} &=& {1\over2}\biggl( \ket0 \otimes \ket0
  + \ket0 \otimes \ket1 \nonumber\\
&& + \ket1 \otimes \ket0
  - \ket1 \otimes \ket1 \biggr) \otimes \ket0\ .
\end{eqnarray}
From a product state, the two dots have evolved to a state of maximal
entanglement.  The full set of transformations are summarized here:
\begin{eqnarray}
\ket0\ket0\ket0  \rightarrow \ket0\ket0\ket0 \rightarrow
  & \ket0\ket0\ket0 & \rightarrow \ket0\ket0\ket0 \nonumber\\
\ket1\ket0\ket0  \rightarrow -i \ket1\ket0\ket1 \rightarrow
  & - i \ket1\ket0\ket1 & \rightarrow \ket1\ket0\ket0 \nonumber\\
\ket0\ket1\ket0 \rightarrow \ket0\ket1\ket0 \rightarrow
  & \ket0\ket1\ket0 & \rightarrow \ket0\ket1\ket0 \nonumber\\
\ket1\ket1\ket0 \rightarrow -i \ket1\ket1\ket1 \rightarrow
  & i \ket1\ket1\ket1 & \rightarrow - \ket1\ket1\ket0
\end{eqnarray}

In practice, noise or decoherence places severe limits on the entanglement
process discussed above.
We can model the effects of noise by replacing Schr\"odinger's equation
with a Markovian master equation for the system,
\begin{equation}
{d\rho\over dt} = - i [\H,\rho] + \sum_k 2 \L_k\rho\Ldag_k
  - \Ldag_k\L_k\rho - \rho\Ldag_k\L_k\ ,
\end{equation}
where $\H$ is the Hamiltonian (\ref{hamiltonian}) above, and the
$L_k$ are a set of operators chosen to model the effects of the noise.
We consider three different cases:  decoherence between $\ket0$ and $\ket1$,
where
\begin{eqnarray}
\L_1 &=& \sqrt\Gamma \ket0\bra0 \otimes \id \otimes \id\ , \\
\L_2 &=& \sqrt\Gamma \id \otimes \ket0\bra0 \otimes \id\ ,
\label{dephasing}
\end{eqnarray}
radiative decay between $\ket1$ and $\ket2$, where
\begin{eqnarray}
\L_1 &=& \sqrt\Gamma \ket1\bra2 \otimes \id \otimes \id\ , \\
\L_2 &=& \sqrt\Gamma \id \otimes \ket1\bra2 \otimes \id\ ,
\label{emission}
\end{eqnarray}
and cavity loss, where
\begin{equation}
\L_1 = \sqrt\Gamma \id \otimes \id \otimes \a\ .
\label{cavity_loss}
\end{equation}
$\Gamma$ represents the noise rate.  The important parameter in assessing
the effects of noise is the size of $\Gamma$ compared with $g$;
if $\Gamma$ is small, we expect entanglement to still be possible.
(Of course, in general, all three types of noise might well be present,
with different $\Gamma$s; we are modeling the case where one form of
noise dominates over the others.)

Figure 3 plots the entanglement of formation
\cite{Wootters1997} as a function
of the relative size of $\Gamma$ and $g$ for the three different
types of noise.  With $\Gamma \ll g$,
substantial entanglement between the
two dots can still be retained.  As shown in Fig. 3,
the entanglement is the least tolerant to decoherence between $\ket0$ and 
$\ket1$ and is the most tolerant to decay from $\ket1$ to $\ket2$,
since the latter decoherence occurs only for the relatively short duration
of the relevant optical transitions.  Since $\ket0$, the lowest excited
state, is also a dark state, a ratio
$\Gamma/g < 10^{-3}$ can be achieved for decoherence 
between $\ket0$ and $\ket1$ if we assume that the decoherence is limited by
radiative decay.  For a $Q$-factor near $10^9$, a ratio
$\Gamma/g$ near $10^{-3}$ is also expected. 

To probe whether a nanocrystal is in the ground or the dark excited state,
one can for example shift the nanocrystal resonance to near a given WGM
resonance and measure the induced change in the resonant frequency of the
WGM.  The WGM resonance remains unchanged when the nanocrystal is in a dark
excited state since the dipole transition is bleached.  Significant change
in the resonant frequency of the WGM results when the nanocrystal remains in
the ground state.  This gives a readout only of the individual qubit
states, which does not allow one to detect entanglement
directly.  However, by examining the statistics of measurement results
as we perform rotations on the internal states of the qubits, it should
be possible to establish coherence; this is analogous to the procedure
used in experiments on entanglement in ion traps \cite{Turchette1998}.


The above 
scheme can be scaled up for quantum information processing.  One
could select and then attach $N$ nanocrystals, each with different but known
resonance
frequencies, to the equator of a single microsphere.  Laser pulses
could be addressed to any of the dots individually, while an applied
voltage could bring any of the dots into resonance with the WGM.  Quantum logic
gates such as the one described above could be applied between any two
dots by exchange of a photon, while laser pulses would enable a large
family of rotations of the internal states of the QDs.  Together, these
two kinds of gates are sufficient to perform quantum computation.  The length
of such computations will again be limited by the inevitable decoherence
processes discussed earlier.  


While the above theoretical analysis indicates that entanglement of
nanocrystals through photon exchange is feasible by using a combined
nanocrystal-microsphere system, the experimental challenge to implement the
proposed scheme is still considerable.  We anticipate the major obstacles are
surface related problems including charge or field fluctuations on the
surface of nanocrystals \cite{Empedocles1997} and relaxation of optical
excitations 
into or via surface states.  Another limiting factor is decoherence due to
electron-phonon
interactions.  In ideal QDs and for low-lying excited 
states, decoherence due to electron-phonon interactions can in principle be
suppressed by strong 3D quantum confinement \cite{Bockelmann1994}.  In this
limit,
radiative decay is the dominant decoherence mechanism at very low
temperature and under low excitation levels.
With continued progress toward fabricating ideal nanocrystals especially in
expitaxial capping of nanocrystals, a microcavity system where nanocrystals
or artificial atoms covalently bounded to a fused silica microsphere should
eventually
provide us a realizable model system for studies of quantum entanglement of
QDs and for implementing at least a modest number of quantum logic gates.
Note that preliminary
studies have already demonstrated extremely high $Q$-factor for combined
nanocrystal-microsphere systems \cite{Fan1999}.

Finally, we note that the combined nanocrystal-microsphere system discussed
above not only takes
advantage of the long-lived lowest excited state in a QD as well as one of
the best available optical microresonators, but also allows
separate fabrication and then assembling of individual QDs and microspheres.
This unique flexibility of ``quantum assembling''
can be an important advantage over epitaxially-grown quantum
dot-microcavity systems.


In conclusion, we suggest that this physical system is a promising
candidate for quantum information processing, and have demonstrated one
possible scheme for producing entanglement.  It may well be possible to
adapt other procedures to this system, such as the recent proposal
of Imamoglu et al. that takes advantage of very long spin decoherence time
\cite{Imamoglu1999}.  Experiments should bring
answers to the questions that remain.


TAB would like to acknowledge helpful conversations with Howard Carmichael,
Oliver Cohen, and Robert Griffiths, and appreciates the hospitality of
the University of Oregon.  TAB was supported in part by NSF grant
No. PHY 96-02084; HW was supported by AFOSR and NSF.

\bigskip

Fig.~1:     Schematics of a combined nanocrystal-microsphere system where
CdS-capped CdSe nanocrystals are attached to the surface of a fused silica
microsphere.
\bigskip

Fig.~2:     Schematics of the energy level structure used.  $\ket1$, the
ground state, and $\ket0$, the lowest excited state and a dark state, serve
as a qubit.  The $\ket1$ to $\ket2$ transition is dipole
allowed.  Higher excitations are neglected.
\bigskip

Fig.~3:  Entanglement of formation vs. noise rate for decoherence
between $\ket0$ and $\ket1$ (solid line), radiative decay from $\ket2$ to
$\ket1$ (dashed line), and cavity loss of the WGM (dotted line).  The noise
rate is given as a fraction of the dipole coupling strength
$g$ between the nanocrystal and the cavity mode.
\bigskip

\begin{figure}[t]
\begin{center}
\epsfig{file=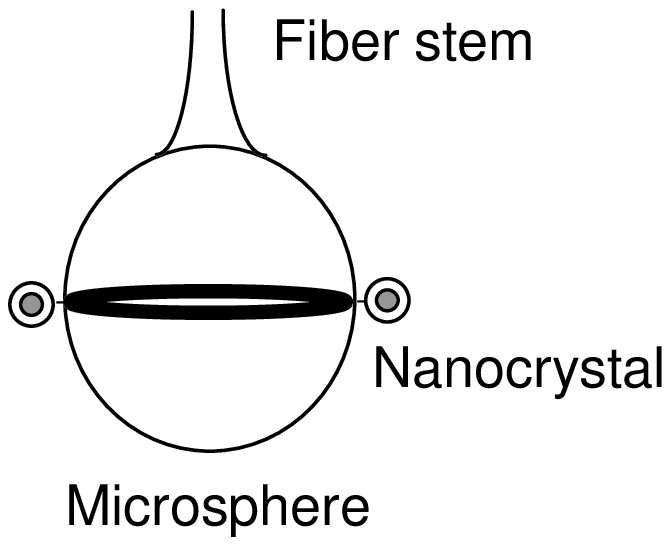, width=5in}
\label{fig1}
\end{center}
\caption{ }
\end{figure}

\begin{figure}[t]
\begin{center}
\epsfig{file=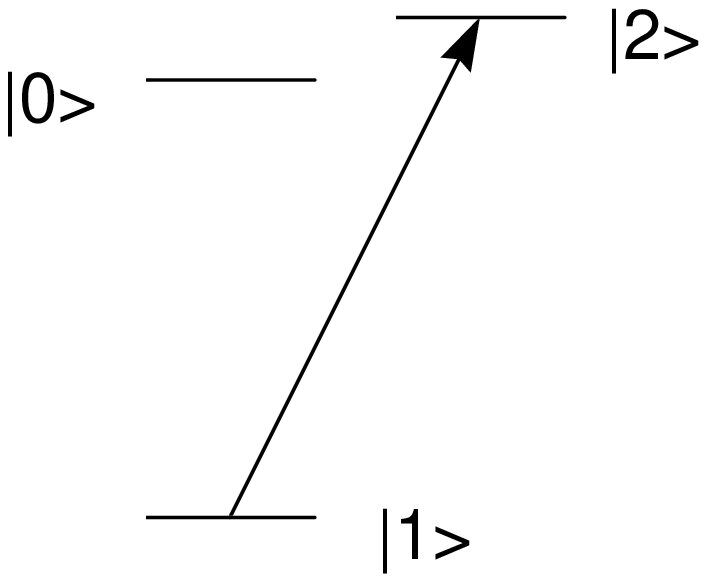, width=5in}
\label{fig2}
\end{center}
\caption{ }
\end{figure}

\begin{figure}[t]
\begin{center}
\epsfig{file=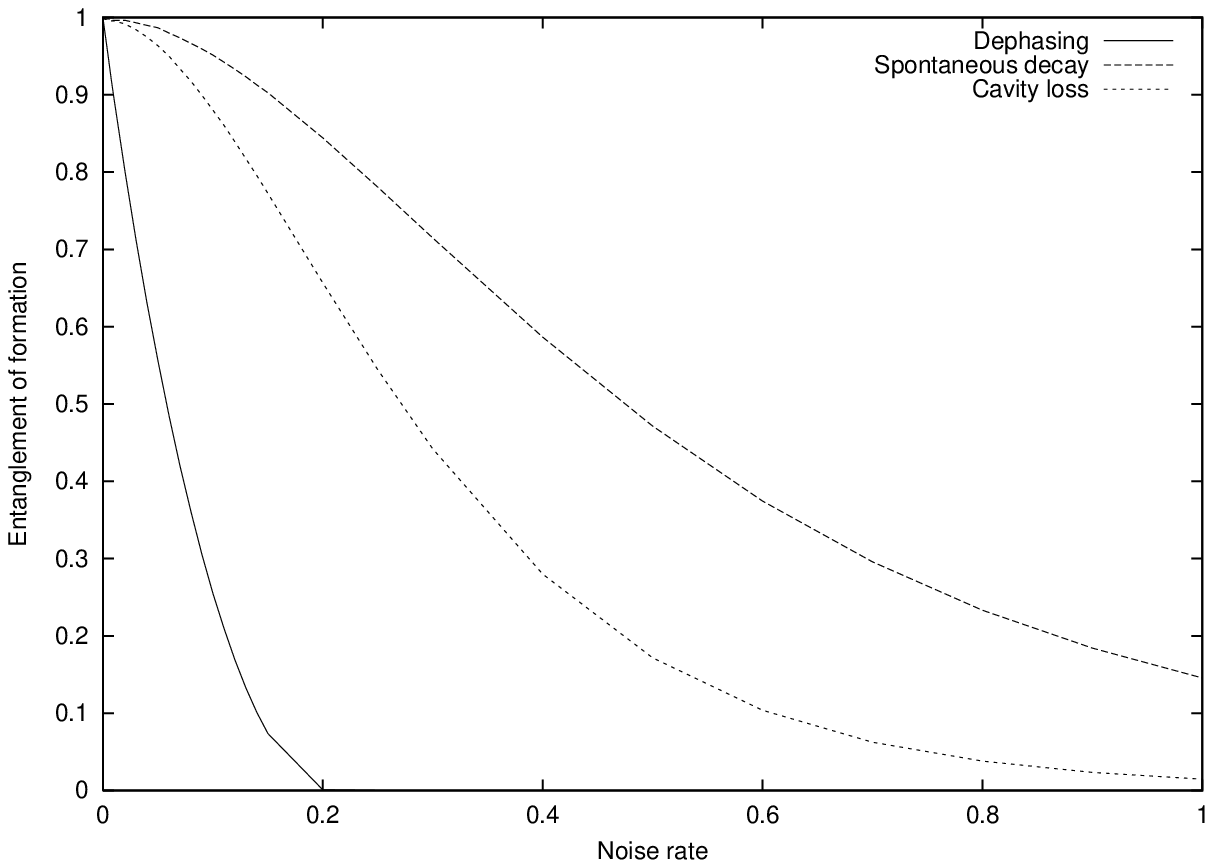, width=3in}
\label{fig3}
\end{center}
\caption{ }
\end{figure}

\end{document}